\documentclass[%
 reprint,
superscriptaddress,
amsmath,amssymb,
aps,
prl,
floatfix
]{revtex4-1}

\usepackage{multirow}
\usepackage{graphicx}
\usepackage{dcolumn}
\usepackage{bm}
\usepackage[version=3]{mhchem}
\usepackage{nicefrac}

\usepackage{color}
\usepackage{dcolumn}
\usepackage{amssymb}
\usepackage{url}
\usepackage{hyperref}
\usepackage{xfrac}
\usepackage{tabularx}
\usepackage{xspace}
\usepackage{amsmath}
\usepackage[normalem]{ulem}
\usepackage{mathtools}
\usepackage[export]{adjustbox}

\def\be{\begin{equation}}
\def\ee{\end{equation}}
\def\bea{\begin{eqnarray}}
\def\eea{\end{eqnarray}}
\def\ba{\begin{array}}
\def\ea{\end{array}}
\def\O{\mathcal{O}}

\newcommand{\PG}[1]{\textcolor{black}{#1}}

\begin{document}
\title{Maple Leaf Antiferromagnet in a Magnetic Field}
\author{Pratyay Ghosh}
\email{pratyay.ghosh@physik.uni-wuerzburg.de}
\affiliation{Institut f\"ur Theoretische Physik und Astrophysik and W\"urzburg-Dresden Cluster of Excellence ct.qmat, Julius-Maximilians-Universit\"at W\"urzburg, Am Hubland Campus S\"ud, W\"urzburg 97074, Germany}
\author{Jannis Seufert}
\affiliation{Institut f\"ur Theoretische Physik und Astrophysik and W\"urzburg-Dresden Cluster of Excellence ct.qmat, Julius-Maximilians-Universit\"at W\"urzburg, Am Hubland Campus S\"ud, W\"urzburg 97074, Germany}
\author{Tobias M\"uller}
\affiliation{Institut f\"ur Theoretische Physik und Astrophysik and W\"urzburg-Dresden Cluster of Excellence ct.qmat, Julius-Maximilians-Universit\"at W\"urzburg, Am Hubland Campus S\"ud, W\"urzburg 97074, Germany}
\author{Fr\'ed\'eric Mila}
\affiliation{Institute of Physics, Ecole Polytechnique F\'ed\'erale de Lausanne (EPFL), CH-1015 Lausanne, Switzerland}
\author{Ronny Thomale}
\email{rthomale@physik.uni-wuerzburg.de}
\affiliation{Institut f\"ur Theoretische Physik und Astrophysik and W\"urzburg-Dresden Cluster of Excellence ct.qmat, Julius-Maximilians-Universit\"at W\"urzburg, Am Hubland Campus S\"ud, W\"urzburg 97074, Germany}

\begin{abstract}
We analyze the quantum antiferromagnet on the maple leaf lattice in the presence of a magnetic field. Starting from its exact dimer ground state and for a magnetic field strength of the order of the local dimer spin exchange coupling, we perform a strong coupling expansion and extract an effective hardcore boson model. The interplay of effective many-body interactions, suppressed single-particle dynamics, and correlated hopping gives way to an intriguing series of superfluid to insulator transitions which correspond to magnetization plateaux in terms of the maple leaf spin degrees of freedom. While we find plateaux at intermediate magnetization to be dominated by bosonic density wave order, we conjecture plateau formation from multi-boson bound states due to correlated hopping for lower magnetization. 
\end{abstract}

\maketitle

\textit{Introduction.} Frustrated quantum antiferromagnets have become a crystallization seed for a plethora of correlated quantum many-body phenomena~\cite{Diepbook,frustrationbook}. Among them, dimer states take a particular role, as they allow for the formulation of a natural local basis encoding quantum entangled degrees of freedom. Singlet dimer states were the first discovered exact ground states of quantum antiferromagnets~\cite{Majumdar1969}, and have taken a pivotal role in connecting frustrated magnetism to contemporary frontiers of condensed matter physics such as fractionalization, topological order, and lattice gauge theories~\cite{Rokhsar1988,Moessner2001,Misguich2002,Wen1991}. A major direction of research on dimer models relates to quantum magnetization plateaux, where a quantum plateau does not originate from semi-classical order by disorder phenomena, but in its essence traces back to quantum entanglement~\cite{Takigawa2011-py}. This hints at the formidable suitability of dimer ground states to address such quantum plateaux, since their quantum nature is already engraved in the very basis of a dimer singlet state such as formed by two adjacent spin-1/2 degrees of freedom. 

\begin{figure}
\includegraphics[width=0.8\columnwidth]{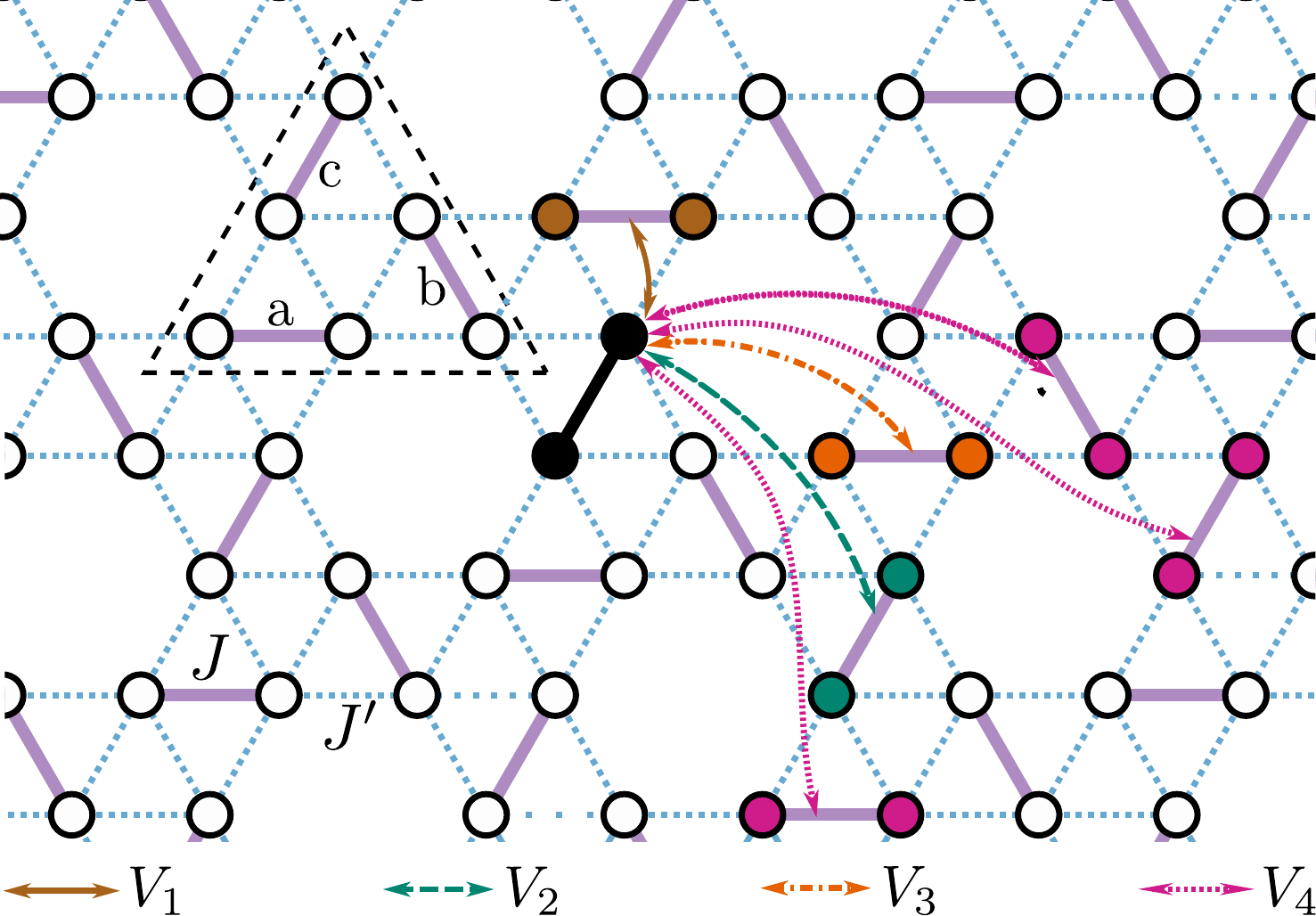}
\caption{Maple leaf model in~\eqref{eq-hamil}. The dotted blue and thick purple bonds relate to an exchange coupling  $J^{\prime}$ and $J$. The 6-site unit cell has three dimer sublattices a, b, and c (dashed triangle frame). The effective triplet-triplet repulsion (reference triplet in black) is expanded into pseudopotentials $V_{1,\dots,4}$~\eqref{eq-int}.  The bonds related to $V_{1,\dots,3}$ ($V_4$) unfold from one (three) bond(s) plus lattice symmetry operations.} \label{fig-1}
\end{figure}

Recently, the quantum antiferromagnet on the maple leaf lattice has been found by three of us to host an exact dimer ground state with exceptional stability~\cite{Ghosh2022}. As compared to the Shastry-Sutherland model (SSM)~\cite{Shastry1981} which has been the dominant cornerstone with an exact dimer ground state for decades\PG{~\cite{SSM_Review,Ghosh2023}}, the Maple Leaf Model (MLM) exhibits a larger dimer ground state domain in terms of minimal bond anisotropy, along with intriguing adjacent magnetic and paramagnetic regimes that are under active investigation~\cite{Ghosh2022,Schmalfuss2002,Farnell2011}.  A particularly interesting domain of a model with a dimer ground state in a magnetic field is reached when the magnetic field strength is of the order of the dimer singlet-triplet gap, i.e., the exchange coupling scale $J$ (Fig.~\ref{fig-1}). At low energies, this then allows one to perform the hardcore boson projection onto a two-dimensional basis formed by the dimer state and the field-aligned triplet state component. The effective hardcore boson model naturally shows a highly exotic arrangement of terms entering the effective Hamiltonian, which are hard to realize in other condensed matter contexts: the single particle dynamics are typically rather suppressed, promoting the relevance of interactions~\cite{Miyahara1999,Miyahara2000}. Depending on the bosonic filling, i.e., the magnetization, and the bond anisotropy, long-range two-body and higher-body interactions arise, combined with correlated hopping terms that can overcome the suppressed single-particle dynamics. There is in principle three major ways for such a bosonic model to develop a phase which, in terms of its underlying spin degrees of freedom, would correspond to a magnetization plateau. First, the system could break translation symmetry and form density waves. Second, the correlated hopping can give rise to bound state formation, which might be particularly relevant for low magnetization~\cite{Momoi2000,Totsuka2001,Dorier2008,Corboz2014,Schneider2016}. Third, topological ordered states of bosons with a finite condensation energy could turn out to be preferred energetically, which would show no sign of translation symmetry breaking~\cite{Misguich2001,Moessner_2010}. This is where a slight analogy to fractional quantum Hall effect can be drawn, where the Wigner crystal state at low densities is replaced by a topologically ordered state at higher densities whose quantum fluctuations reinstall translation symmetry.

In this Letter, we analyze the MLM in a magnetic field. The Hamiltonian is given by
\be\label{eq-hamil}
\mathcal{H}=\mathcal{H}_{0}+\mathcal{H}'-B \sum_{i} S_{i}^{z}
\ee
with
$$
\mathcal{H}_{0} =J \sum_{\langle i j\rangle} \vec{S}_{i} \cdot \vec{S}_{j} \text{, and } 
\mathcal{H}' =J^{\prime} \sum_{\langle k l\rangle} \vec{S}_{k} \cdot \vec{S}_{l},
$$
where $\vec{S}_{i}$ denotes a spin vector operator acting on a spin-$1/2$ degree of freedom at site $i$. Aside from a Zeeman term, there are two separate summations $i j$, and $k l$ over the nearest neighbor bonds denoted purple (thick) and blue (dotted), respectively (Fig. \ref{fig-1}). Upon a hardcore boson projection for $B\sim J$, we perform a perturbative expansion in the non-dimer exchange bond strength $J'$. In terms of single-boson self-energy expansion, we find the single-particle dynamics to cancel up to ninth order in $J'/J$, hinting at the exceptional suppression of triplet kinetic energy in the MLM. We extract the triplet-triplet repulsive interactions up to fourth order in $J'$, and perform Monte Carlo simulations\PG{(see Supplemental Material ~\cite{supp} which also contains Refs.~\cite{Betts1995,Gelfand-1991,Caspers_1984,Chung2001,deconfined,Mambrini2006})} to obtain density wave type plateaux at $m=1/6$, $2/9$, $2/7$, and $1/3$. We identify the formation of two-particle and, possibly, three-particle bound states from correlated hopping which may lead to additional plateaux at lower magnetization. 

\textit{Spin gap and single-particle dynamics.} 
The ground state of $\mathcal{H}_{0}$ is a product state of spin singlets on all the purple bonds for $J'/J\lesssim0.74$~\cite{Ghosh2022}. The eigenstates of an isolated purple dimer are the singlet state $|s\rangle$ and the three triplet states $|t_{1}\rangle, |t_{0}\rangle$, and $|t_{-1}\rangle$, where the subscript denotes the total $S^{z}$. The effect of $\mathcal{H}'$ is included perturbatively in the basis of the dimer eigenstates~\cite{Sakurai2020}. The spin, i.e., singlet-triplet gap reads
\be\label{eq-delta}
\Delta = J-B-\frac{J'^2}{J}-\frac{J'^{3}}{2 J^{2}}\PG{-\frac{5J'^4}{8J^3}}+\O(J'^5).
\ee 
The expression of $\Delta$ in MLM is identical to the SSM up to \PG{third} order~\cite{Miyahara1999}. 

\begin{figure}
\includegraphics[width=0.95\columnwidth]{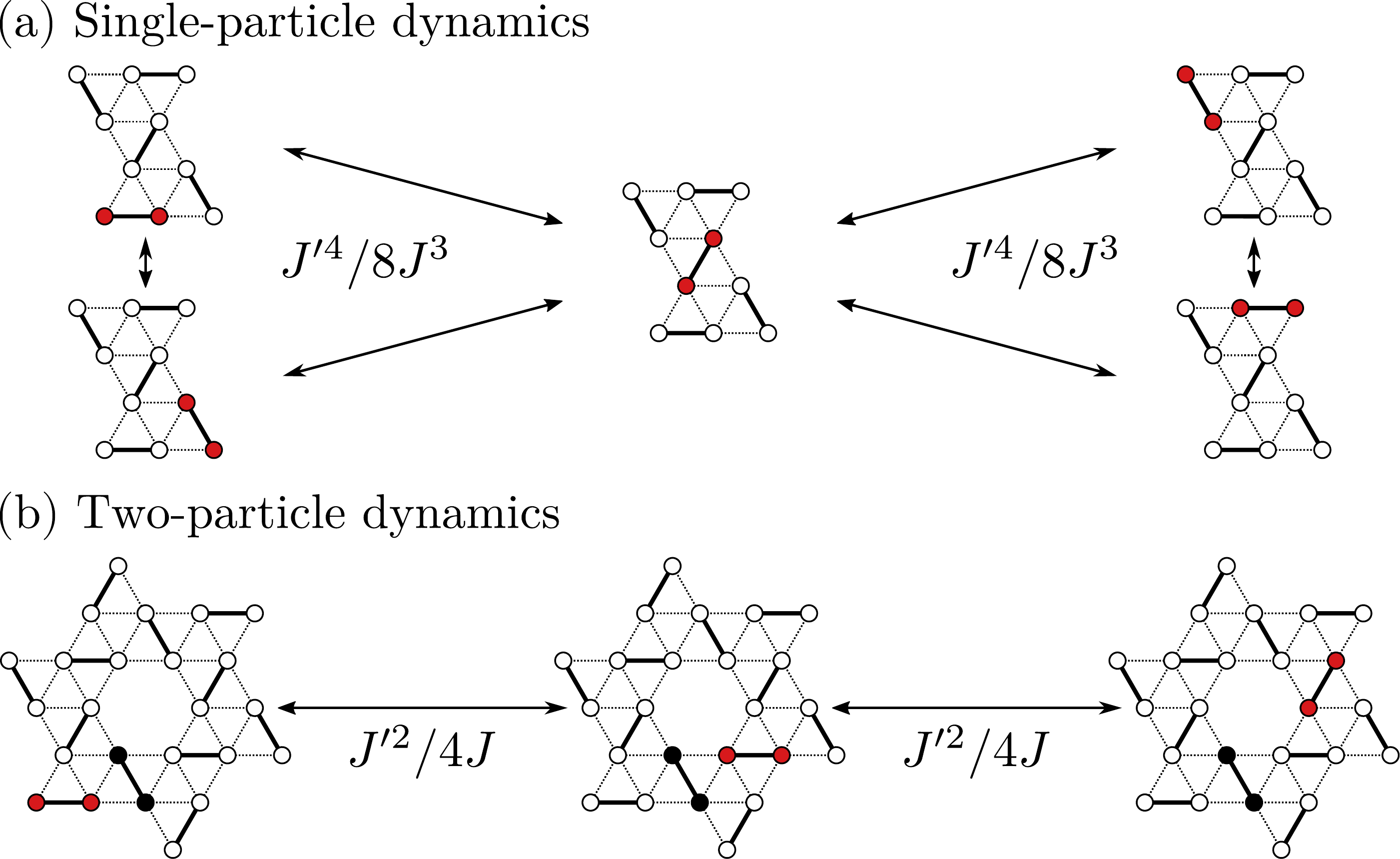}
\caption{(a) Hopping of a single triplet (marked by the dimer filled in red) at \PG{fourth}-order in $J'$. There are two symmetry-inequivalent ways a triplet can move between two equal-sublattice dimers. (b) For correlated hopping, one triplet (filled in red) hops assisted by another triplet in its vicinity (filled in black). Such a hopping emerges from second order in $J'$ and implies triplet bound states.} \label{fig-2}
\end{figure}

The parity of the singlets and the matrix elements of $\mathcal{H}'$ impose strong constraints on the hopping processes of the triplets, namely, a triplet can only move to two of its four neighboring dimers, and when it does, it leaves another triplet behind. Consequently, we see that the hopping of single triplets starts in \PG{fourth-order} perturbation theory, where a triplet can \PG{hop to any of the neighbouring dimers}(Fig.~\ref{fig-2}(a))~\cite{supp}. For SSM, such a hopping  appears only at sixth-order~\cite{Miyahara1999}. The \PG{enhancement} of triplet hopping for MLM is rooted in its lattice geometry~\cite{supp}. 

\textit{Two-particle dynamics.}
An important property of the dimer model is that it allows nontrivial two-triplet hopping called \emph{correlated hopping}, where one triplet can hop assisted by a second triplet in its proximity. Note that a process like that can neither be interpreted as single-triplet hopping nor a triplet pair creation-annihilation. Correlated hopping motion of two triplets occurs from a second order perturbation  (Fig.~\ref{fig-2}(b))~\cite{supp}.  Since the \PG{single triplet excitations occur at further two-orders of perturbation}, the correlated hopping plays a crucial role. That has likewise been noticed in the SSM, where the correlated hopping results in two-triplets bound states~\cite{Momoi2000,Totsuka2001,Dorier2008,Corboz2014,Schneider2016}. 

\textit{Hardcore boson projection.} 
For $B>0$, the lowest energy states with magnetization $m \geq 0$ for the whole system consist of $|s\rangle$ and $|t_{1}\rangle$, which we project to as the low-energy physical degrees of freedom. Considering $|t_{1}\rangle$ as a hard-core boson and $|s\rangle$ as a vacancy, we project our system onto an effective hard-core boson model along with a perturbative treatment around the limit $J'/J\ll1$. Whereas the $|t_1\rangle$ triplets are treated as on-shell magnetic particles, the rest of the dimer states, i.e., $|t_0\rangle$ and $|t_{-1}\rangle$, are considered as intermediate virtual states in perturbation theory. We estimate the interaction energy between two hard-core bosons for which we calculate the energy required to create two triplet excitations from the ground state as a sum of (i) the spin gap energy required to generate a triplet excitation ($2\Delta$), which includes the self-energy-type corrections outlined in \eqref{eq-delta}, and (ii) the interaction energy between the excited triplets which we call pseudopotentials (PPs) $V_n$. The $n$ in the subscript refers to the fact that two triplets can interact in different orders in perturbation theory due to their different relative positions. In Fig.~\ref{fig-1}, we depict the interactions (symmetry reduced) between different triplet pairs that interact within third-order perturbation theory. We thus find the PPs
\begin{equation}
\begin{aligned}
V_{1} &= \frac{J'}{2}+\frac{J'^{2}}{2 J}+\O(J'^4) \\
V_{2} &= \frac{J'^{2}}{2 J}+\frac{3 J'^{3}}{4 J^{2}}+\O(J'^4) \\
V_{3} &= \frac{J'^{3}}{8 J^{2}}+\O(J'^4) \label{eq-int}
\end{aligned}
\end{equation}
Clearly, the triplet-triplet interactions do not truncate at this order, but the higher order calculations are analytically rather cumbersome. This is where extensive numerical methods, e.g. perturbative continuous unitary transformations~\cite{Dorier2008}, are ideal to take over, which we keep as a future endeavor.

\textit{Density wave magnetization plateaux.}
\begin{figure}
\includegraphics[width=0.8\columnwidth]{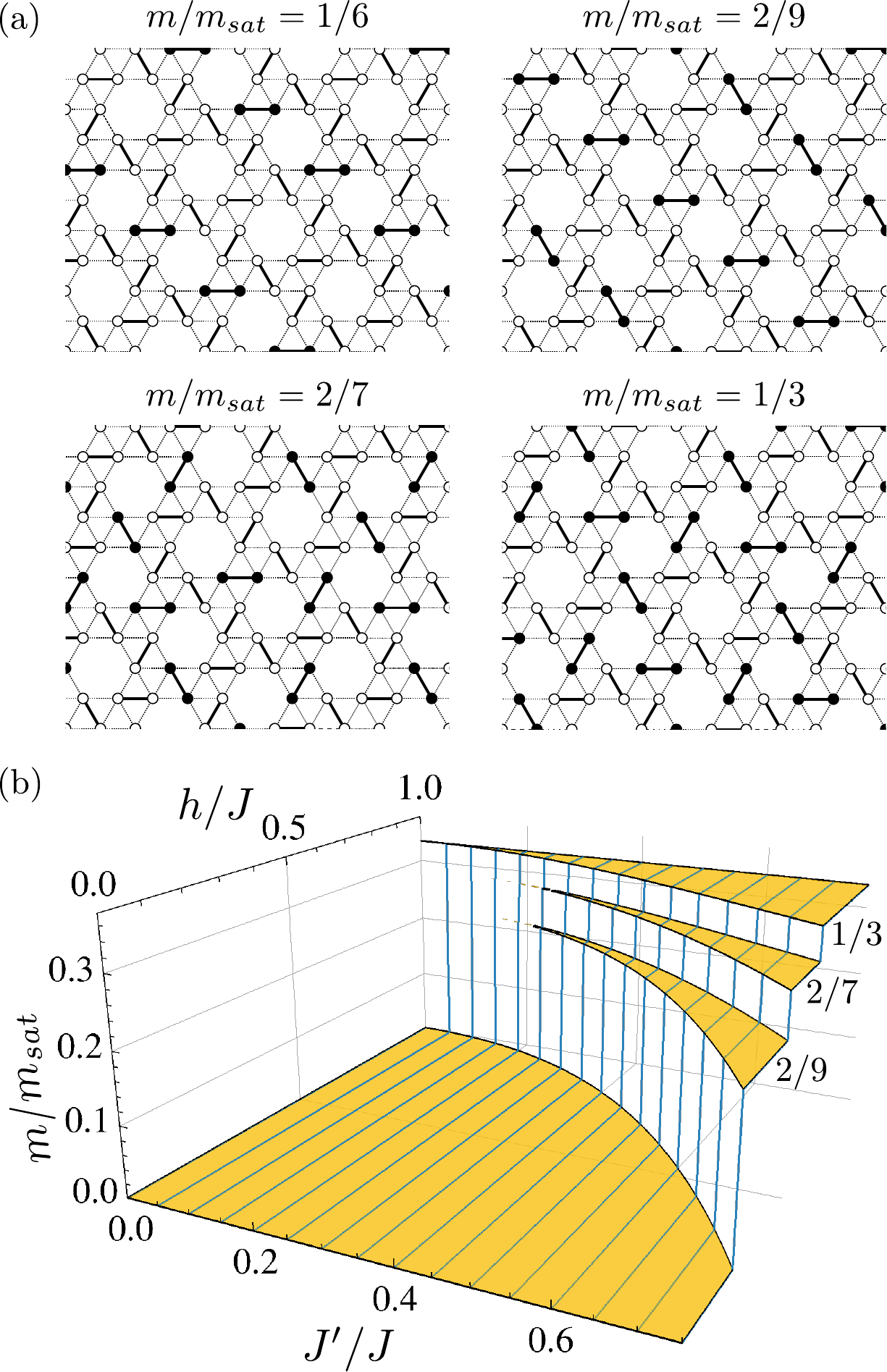}
\caption{(a) The density wave structures at magnetization plateaux. The triplet excitations (singlets) are shown by filled (open) dimers. (b) Magnetization processes of the MLM as a function of $J^\prime$ and $h$ upto third order in perturnation.} \label{fig-3}
\end{figure}
We first discuss the magnetization plateaux of~\eqref{eq-hamil} through Monte Carlo simulations\PG{~\cite{supp}} of the effective hard-core boson model, with PPs defined in~\eqref{eq-int}, in the limit of negligible kinetic energy. We determine magnetization curves \PG{upto third-order in perturbation} for different values of $J'/J$ which yield plateaux at $m/m_{sat}=2/9$, $2/7$, and $1/3$ at sufficiently low temperatures, where $m_{sat}$ denotes the saturation magnetization of full triplet occupancy. The $1/3$ plateau has also been predicted via exact diagonalization~\cite{Farnell2011}. These plateaux only appear when the spin gap $\Delta$ is closed by $B$. The strong repulsive interactions, however, will not allow for all the singlets to become triplets upon gap closing. Due to the competition between the energy gained by triplet excitations and the repulsive energy, one would expect that for a given $B$ the triplets would assume a particular superstructure (density wave) to minimize the energy. Such a density wave with a fixed magnetization is a stable configuration for a finite range in $B$, long-range correlated, and exhibits a spin gap. We construct all possible density waves of hardcore bosons with different unit cells, and identify the particular arrangement that minimizes the energy for a given field. This approach turns out to agree well with our Monte Carlo results. The various bosonic density waves and their corresponding magnetization phase diagram at $T=0$ are presented in Fig.~\ref{fig-3}. \PG{The inclusion of higher-order calculations, however, would produce PPs beyond the ones mentioned in \eqref{eq-int} and can produce plateaux with lower magnetization. For example, we depict the PPs that would appear in fourth-order perturbation theory, \begin{equation}V_4=\frac{J'^{4}}{8 J^{3}}+\O(J'^5),\end{equation} in Fig.~\ref{fig-1}. The inclusion of $V_4$ would introduce a $m/m_{sat}=1/6$ plateau (see superstructure in Fig.~\ref{fig-3}).}  
We, also, do not explore the magnetization beyond the $1/3$ plateau, as from there on we would encounter more than one hard-core boson per unit cell, and hence three-particle interactions would need to be considered to properly investigate such plateaux of higher magnetization.  

While the Monte Carlo simulations of the hardcore boson projection is suitable to capture the static physics of the density wave magnetization plateaux, understanding how the system rearranges itself across a critical field requires the inclusion of correlated hopping. When a plateau state is destabilized by a field, the system would still maintain the density wave of the preceding phase, but would have extra bosons in the system. These bosons can perform superfluid motion in the lattice through correlated hoppings assisted by the bosonic density wave. Thus, as a combination of the density wave and the superfluid, the resulting state near criticality is a supersolid phase~\cite{Matsuda2013,Shi2022}, where the magnetization should increase smoothly with the increasing field and finally reach the next plateau state. In addition to the plateau states depicted in Fig.~\ref{fig-3}, we thus also conjecture supersolid phases to appear near the plateau transitions. Not all of these supersolid phases, however, should a priori be expected to be stable~\cite{Momoi2000,Gan_2011}. 
     
\textit{Triplet bound states.}
\begin{figure*}
\includegraphics[width=0.9\textwidth]{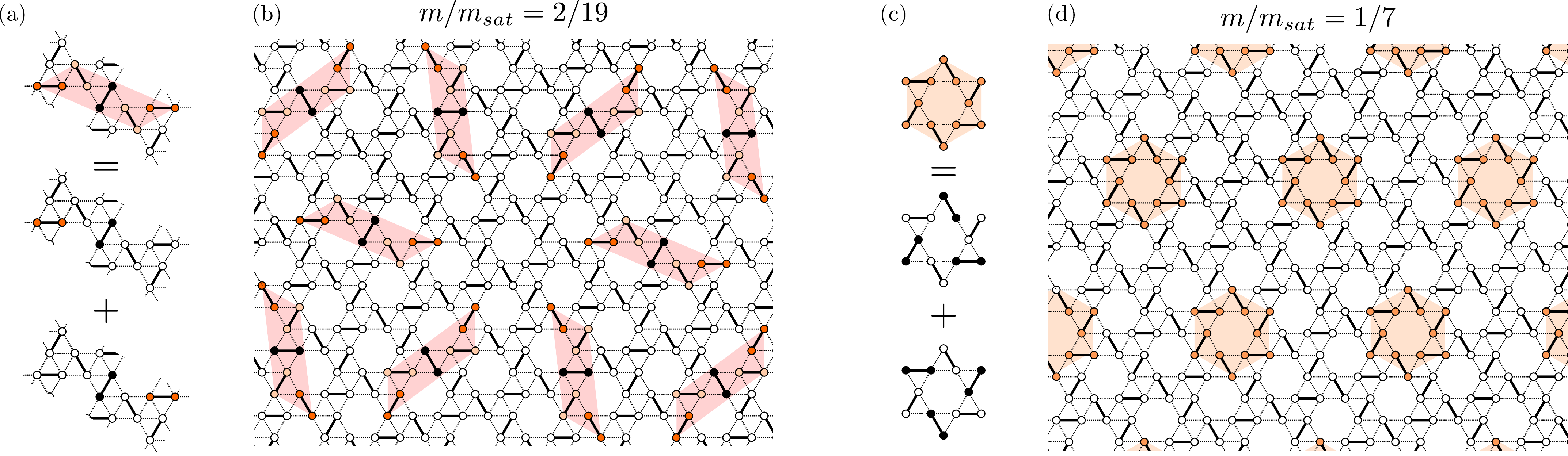}
\caption{(a) Two- and (c) possible three-particle bound states, and the constituent states with dominant contributions. The filled dimer end-points reprents a triplet. The light red paralellograms in (a) and (b), and light orange hexagons in (c) and (d) represent the bound states of two and three triplets, respectively. Speculative crystal of (b) two- and (d) three-particle bound states giving rise to $m/m_{sat}=2/19$ and $1/7$. The bound states of two and three triplets are arranged to interact only via $V_4$. } \label{fig-4}
\end{figure*}
When two excited triplets are far apart from each other, individual triplets are nearly localized and gain only minuscule energy from kinematics. When they are adjacent to each other, however, correlated hopping makes the coherent motion of two triplets possible, and hence the gain of kinetic energy (Fig.~\ref{fig-2}(b)). The balance between PPs and correlated hopping can give rise to $S^z=2$ bound states, i.e. a bound state of two on-shell $|t_1\rangle$ triplets: When we place two triplets next to each other, the interaction between them is linear in $J'$ and typically dominant. They can, however, employ correlated hopping $\propto J^{\prime 2}$ and move away from each other in such a way that their PP energy penalty reduces to $V_3$, which is cubic in $J'$, and thus as a whole form a two-particle bound state.

The hopping processes of two triplets can be decomposed into the center-of-mass motion and the relative motion, which leaves us with twelve unique states with different two-triplet configurations. Note that, for SSM, there are only four such states, as the effective lattice of dimer bonds is a square lattice. Applying the same rationale to MLM, the effective dimer lattice is of kagom\'e type, which enhances the complexity of the problem. Upon calculating the energies $e_{2P}$ of a two-triplet excitation~\cite{supp}, we find the energetically favoured two-triplet bound state, which locates in the light red region shown in Fig.~\ref{fig-4} (a), and is mostly composed of a linear combination of two triplet states interacting via $V_3$. The binding energy of this bound state, $e_{2P}-2\Delta$, is found to be $-3J^{\prime4}/16J^3$ (in lowest order) per particle~\cite{supp}. Because the two-particle bound states thus have lower energy than two separate triplets, the gap of the bound states collapses faster than the singlet-triplet gap upon increasing $B$, thus enforcing the bound states to condense first rather than the isolated single triplets. 

For SSM, it is shown that the $S^z=2$ bound states give rise to additional magnetization plateaux near $m=0$ by forming crystals of bound states~\cite{Corboz2014,Schneider2016}. The same phenomenon should be expected for MLM. In Fig.~\ref{fig-4} (b), we speculate on the existence of such a bound state crystal corresponding to  $m/m_{sat}=2/19$. Moreover, in the $2/7$ and $1/3$ plateaux depicted in Fig.~\ref{fig-3} (a), the relative position of the triplets favors the onset of correlated hopping. Therefore, upon inclusion of dynamics, these are likely to gain stability from becoming a crystal of bound states. The $1/6$ and $2/9$ plateaux, on the other hand, can be destabilized by other competing bound state crystals. The small size of the binding energy further suggests that these intricacies are likely to only appear for larger values of $J'/J$.       

There is, however, much more to explore with regard to bound states in the MLM. Drawing a comparison to SSM, the lowest-energy bound state is formed around a square on the effective lattice\PG{~\cite{Corboz2013}}. In our case, the effective lattice is made of triangles and hexagons. No two-particle bound state can be formed around a triangle, as the interaction between the triplet would always be $V_1$, and not around hexagons either due to the range of correlated hopping. Therefore, it is quite natural to expect the appearance of three-triplet bound states, where three particles themselves perform a coherent motion around a hexagon and overcome the mutual repulsion~\cite{supp}. We investigate this regime within a truncated three-particle basis. The binding energy corresponding to the lowest energy state, which is predominantly a linear combination of  the three particle states in Fig.~\ref{fig-4} (c), is found to be $J^{\prime3}/24J^2-J^{\prime4}/8J^3+\O(J'^5)$ per particle, where the binding energy is positive for small $J^\prime$. A more elaborate analysis is required to capture the stability of such three-particle bound states. Under the assumption that such a bound state is stable, we present a speculative  bound state crystal in Fig.~\ref{fig-4} (d) producing a plateau at $m/m_{sat}=1/7$.

\textit{Conclusions and Outlook.} 
We have investigated the magnetization phase diagram of the MLM by using a strong-coupling expansion of a hard-core boson projected effective model. We discover that finite single-particle dynamics only occur by tenth order perturbation theory in $J'$, suggesting a remarkable flatness of the single triplet dispersion. We extract the triplet-triplet repulsive PPs  up to fourth order in $J'$, and find density wave type plateaux at $m/m_{sat}=1/6$, $2/9$, $2/7$, and $1/3$. We predict the appearance of supersolid phases within this unfolding magnetization phase diagram. Furthermore, we emphasize the phenomenological step towards the development of two- and perhaps three-particle bound states from correlated hopping, which might result in an amended stability of density wave magnetization plateaux, and further plateaux at lower magnetization. 

Several pressing issues about the MLM in a magnetic field remain unanswered and require further investigation. First, we truncate our evaluation of two-particle PPs at fourth-order despite that higher-order expansions will produce longer-range repulsion between particles and may induce other lower magnetization plateaux. There, another significant open question is to investigate the crystallization of two-particle, or even three-body, bound states at lower magnetic fields, and the impact of three-body interactions at higher magnetization.

\textit{Acknowledgments.}  \PG{We thank Kai Phillip Schmidt for pointing out a fourh-order hopping process for the triplets that was missed in an early version of the manuscript.} The work in W\"urzburg is supported by the Deutsche Forschungsgemeinschaft (DFG, German Research Foundation) through Project-ID 258499086-SFB 1170 and the Würzburg-Dresden Cluster of Excellence on Complexity and Topology in Quantum Matter – ct.qmat Project-ID 390858490-EXC 2147. \PG{FM has been supported by the Swiss National Science Foundation, Grant No, 182179.} PG and JS contributed equally in this work.

%

\end{document}


\title{Supplementary Materials: Maple Leaf Antiferromagnet in a Magnetic Field}
\author{Pratyay Ghosh}
\email{pratyay.ghosh@physik.uni-wuerzburg.de}
\affiliation{Institut f\"ur Theoretische Physik und Astrophysik and W\"urzburg-Dresden Cluster of Excellence ct.qmat, Julius-Maximilians-Universit\"at W\"urzburg, Am Hubland Campus S\"ud, W\"urzburg 97074, Germany}
\author{Jannis Seufert}
\affiliation{Institut f\"ur Theoretische Physik und Astrophysik and W\"urzburg-Dresden Cluster of Excellence ct.qmat, Julius-Maximilians-Universit\"at W\"urzburg, Am Hubland Campus S\"ud, W\"urzburg 97074, Germany}
\author{Tobias M\"uller}
\affiliation{Institut f\"ur Theoretische Physik und Astrophysik and W\"urzburg-Dresden Cluster of Excellence ct.qmat, Julius-Maximilians-Universit\"at W\"urzburg, Am Hubland Campus S\"ud, W\"urzburg 97074, Germany}
\author{Fr\'ed\'eric Mila}
\affiliation{Institute of Physics, Ecole Polytechnique F\'ed\'erale de Lausanne (EPFL), CH-1015 Lausanne, Switzerland}
\author{Ronny Thomale}
\email{rthomale@physik.uni-wuerzburg.de}
\affiliation{Institut f\"ur Theoretische Physik und Astrophysik and W\"urzburg-Dresden Cluster of Excellence ct.qmat, Julius-Maximilians-Universit\"at W\"urzburg, Am Hubland Campus S\"ud, W\"urzburg 97074, Germany}

\maketitle
\beginsupplement

\section{Model}
\begin{figure}
\includegraphics[width=0.4\columnwidth]{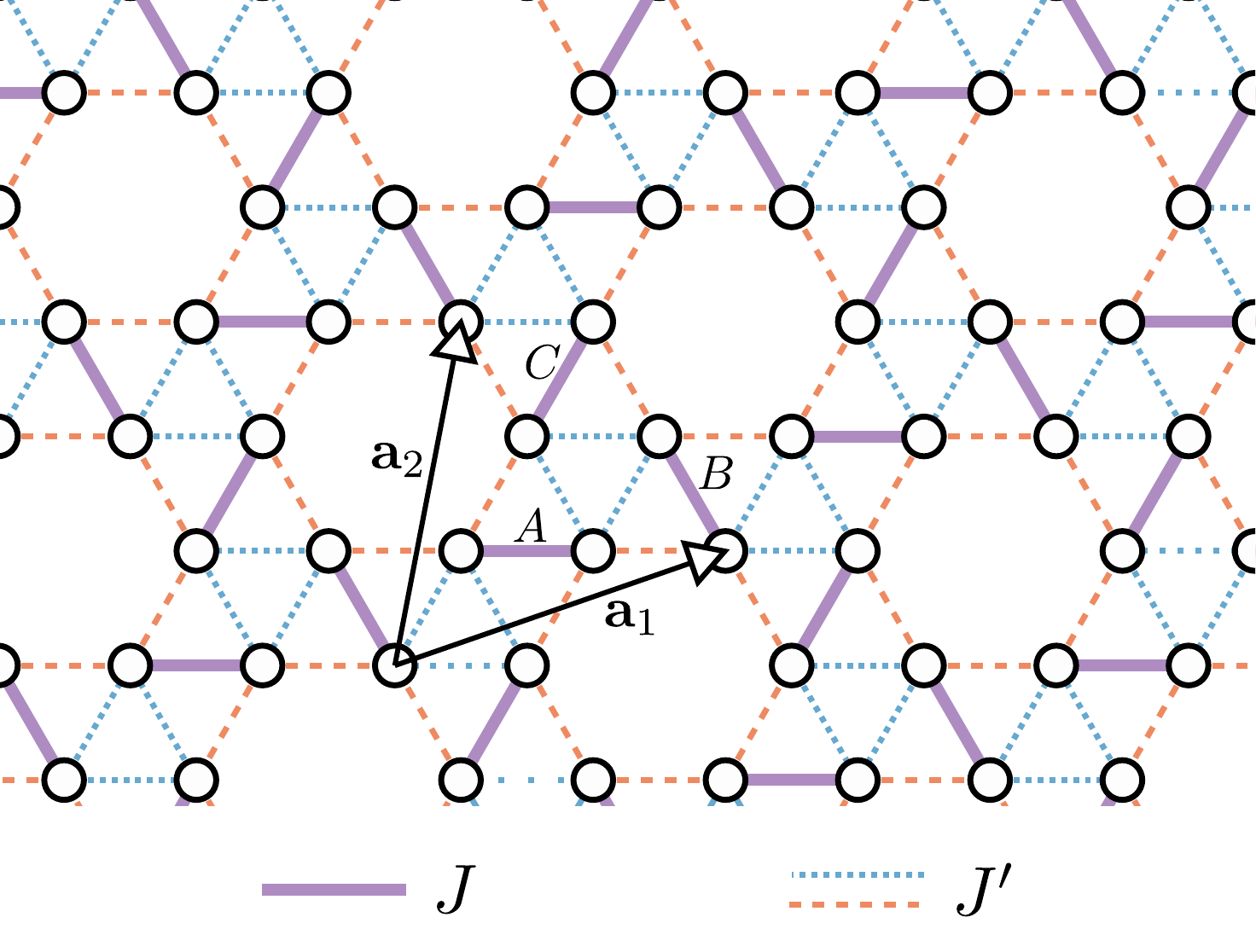}
\caption{Maple-leaf lattice with three symmetry non-equivalent nearest-neighbor bonds. The dashed orange and dotted blue bonds have a Heisenberg exchange with an identical strength of $J^{\prime}$, whereas the purple bonds have a Heisenberg interaction strength of $J$.
It has been demonstrated in \cite{Ghosh2022} that the system has an exactly solvable exact ground state for $J^{\prime}/J\le 0.5$, which is a product state of spin-singlets formed on the $J$-bonds. The lattice vectors are given by $\mathbf{a}_1=\sqrt{7}\hat{x}$ and $\mathbf{a}_1=\sqrt{7}/2(\hat{x}+\sqrt{3}\hat{y})$.} \label{fig-S1}
\end{figure}

The maple leaf is a lattice of triangular motifs, that corresponds to a 1/7-site (1/6-bond) depleted triangular lattice with coordination number 5~\cite{Betts1995}. The lattice has a $p6$ symmetry with three symmetry inequivalent nearest-neighbor bonds (see Fig.~\ref{fig-S1}). The maple-leaf model (MLM) is defined by
\be\label{eq-hamil}
\mathcal{H}=\mathcal{H}_{0}+\mathcal{H}'-B \sum_{i} S_{i}^{z}
\ee
with
$$
\begin{aligned}
\mathcal{H}_{0} &=J \sum_{\langle l m\rangle} \vec{S}_{l} \cdot \vec{S}_{m} \\
\mathcal{H}' &=J^{\prime} \sum_{\langle k l\rangle} \vec{S}_{k} \cdot \vec{S}_{l}+J^{\prime} \sum_{\langle k m\rangle} \vec{S}_{k} \cdot \vec{S}_{m}
\end{aligned}
$$
where, $\vec{S}_{i}$ are spin-$1 / 2$ operators. In addition to the Zeeman term, the separate sums over the three nonequivalent nearest neighbor bonds $l m, k l$, and $k m$ on the maple leaf lattice are denoted in purple (thick), orange (dashed), and blue (dotted), respectively (Fig.~\ref{fig-S1}). The analytic calculation of Ref.~\cite{Ghosh2022} shows that MLM has an exactly solvable exact dimer ground state, a product state of singlets on all purple bonds, for $J'/J\le 0.5$. Further numerical results estimate that this dimer state is the ground state of the system even for $J'/J\lesssim0.74$~\cite{Ghosh2022,Farnell2011}.
\begin{figure}
\includegraphics[width=0.2\columnwidth]{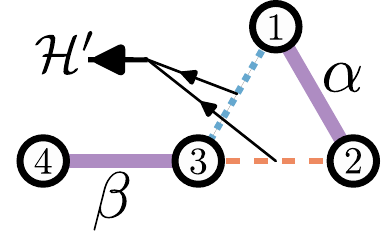}
\caption{Elementary unit for the interaction between a pair of nearest-neighbor dimers.} \label{fig-perturb}
\end{figure}
\section{Matrix Elements of $\mathcal{H}'$}
Starting from the $J>>J'$, for a $B\sim J$, to perform a strong coupling expansion, we, first, move to the dimer basis, which consists of the singlet, $\ket{s}$, and the three triplets, $\ket{t_1}$, $\ket{t_0}$, and $\ket{t_\onebar}$ ($\onebar\equiv-1$), which are given by
\begin{align}
\ket{s} &= \frac{1}{\sqrt{2}}(\ket{\uparrow_1\downarrow_2} - \ket{\downarrow_1 \uparrow_2}) &&
\ket{t_1} = \ket{\uparrow_1\uparrow_2} &&
\ket{t_0} = \frac{1}{\sqrt{2}}(\ket{\uparrow_1\downarrow_2} + \ket{\uparrow_1\downarrow_2}) &&
\ket{t_\onebar} = \ket{\downarrow_1\downarrow_2}
\end{align}
where $1$ and $2$ indicate the site across a dimer. We calculate the matrix elements of $\mathcal{H}'=J^\prime\vec{S}_3\cdot(\vec{S}_1+\vec{S}_2)$ (see Fig. \ref{fig-perturb}) for all the states on any pair of the nearest-neighbor dimer bonds to find
\begin{align}\label{mat-element}
&\mathcal{H}' \ketab{s}{s} =~ 0 \\
&\mathcal{H}' \ketab{s}{t_0} =~ 0 \\
&\mathcal{H}' \ketab{s}{t_m} =~ 0 \\
&\mathcal{H}' \ketab{t_0}{s} = \phantom{\pm}\frac{J^\prime}{2} (\ketab{t_1}{t_\onebar} - \ketab{t_\onebar}{t_1}) \\
&\mathcal{H}' \ketab{t_0}{t_0} = \phantom{\pm}\frac{J^\prime}{2} (\ketab{t_1}{t_\onebar} + \ketab{t_\onebar}{t_1}) \\
&\mathcal{H}' \ketab{t_m}{s} = m \frac{J^\prime}{2} (\ketab{t_m}{t_0} - \ketab{t_0}{t_m})\\
&\mathcal{H}' \ketab{t_m}{t_{\bar{m}}} =~ \frac{J^\prime}{2} (\ket{t_0}_\alpha\ket{t_0}_\beta +m \ket{t_0}_\alpha\ket{s}_\beta - \ketab{t_m}{t_{\overline{m}}}) \\
&\mathcal{H}' \ketab{t_m}{t_m} =~ \frac{J^\prime}{2} \ketab{t_m}{t_m} \\
&\mathcal{H}' \ketab{t_m}{t_0} =~ \frac{J^\prime}{2} (\ketab{t_0}{t_m} +m \ketab{t_m}{s}) \\
&\mathcal{H}' \ketab{t_0}{t_m} =~ \frac{J^\prime}{2} (\ket{t_m}_\alpha\ket{t_0}_\beta +{\bar{m}} \ket{t_m}_\alpha\ket{s}_\beta)
\end{align}
with $m=\pm1$
Here, $(\alpha,\beta)=(A,C)$, $(C,B)$ or $(B,A)$, where $A$, $B$, and $C$ are the three dimer sublattices (see Fig.\ref{fig-S1}).

\section{Single Triplet Dispersion and excitation gap}
\begin{figure*}
\includegraphics[width=\textwidth]{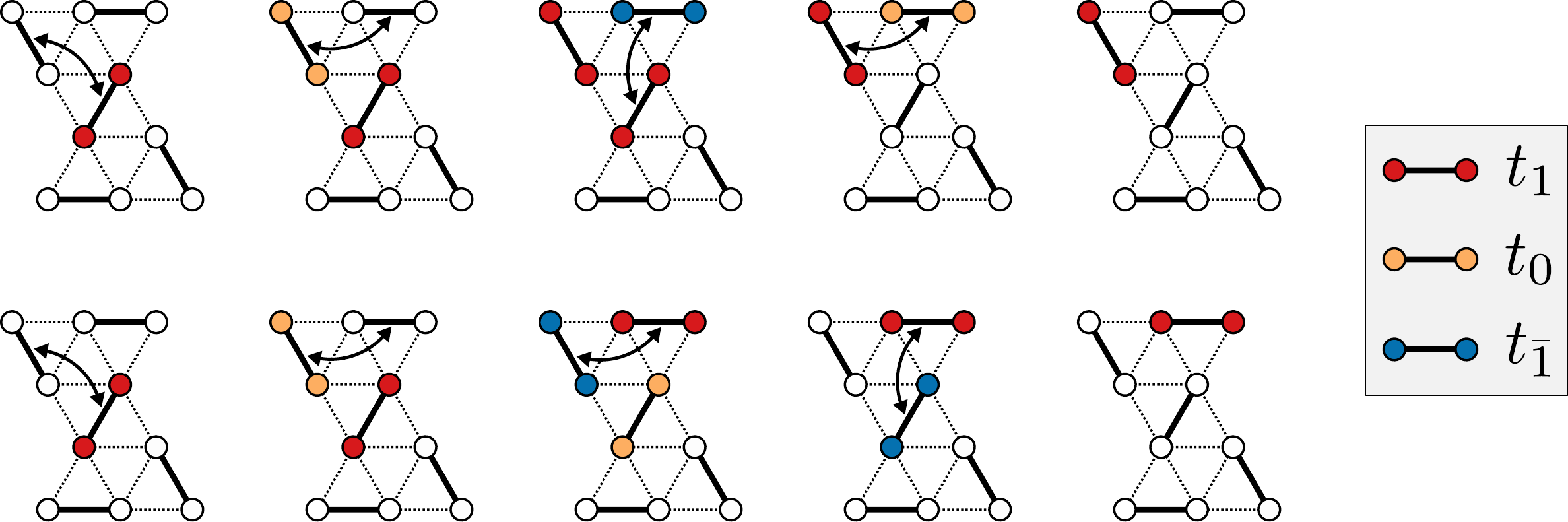}
\caption{Representatives hopping processes of a single triplet excitation. The hopping is only possible from the fourth-order perturbations.} \label{fig-1p_hopping}
\end{figure*}

With energy $e_0=-3 J / 4$ per dimer, the ground state of $\mathcal{H}_0$ is a product state of singlets on each purple bond. $\mathcal{H}'$ in \eqref{eq-hamil} is treated using the standard perturbation theory~\cite{Sakurai2020}. Note that, the singlets possess an odd parity which one should be mindful of while performing the calculations. First, we focus on the hopping processes involving an isolated triplet. After carefully analyzing all the matrix elements, we discover that no triplet hopping is conceivable up to third-order in perturbation -- the hopping only survives at the fourth-order perturbation. Examining Eqs.~\eqref{mat-element} makes the cause behind this suppression of single particle dynamics quite apparent. To display an example, we start from the state with a $t_{1}$ triplet on one $B$-dimer. Now, evidently, $\mathcal{H}'\ket{s}_C\ket{t_1}_B=0$. The states with finite matrix elements with our starting state can be found on the right side of the equation
$$\mathcal{H}'\ket{t_1}_B\ket{s}_A= \frac{J^\prime}{2} \left(\ket{t_1}_B\ket{t_0}_A - \ket{t_0}_B\ket{t_1}_A\right).$$
Interestingly, in this process one triplet state can only result in two triplet states only on the $A$ and $B$ dimer, the $C$ dimer remains inaccessible when $\mathcal{H}'$ is applied only once. These limit a triplet's ability to move independently and only let the triplets hopping possible via a closed path of dimer bonds. These conditions allow the hopping processes only start from the fourth-order in the perturbation. See the two examples of such hopping processes in Fig.~\ref{fig-1p_hopping}. As a comparison, in the one-dimension counterpart of the MLM, the tetrahedral cluster spin chain, the triplet excitations are fully localized~\cite{Gelfand-1991}. For SSM, the same hopping happens in sixth order in perturbation~\cite{Miyahara1999,Miyahara2000,SSM_Review}. The situation is quite different for Majumdar-Ghosh model~\cite{Majumdar1969} where the degeneracy of the singlet ground state allows for a dispersive single spin excitation~\cite{Caspers_1984}.

Starting from the product singlet state, and for $B> 0$, we use the lowest energy $ \ket{s} $ and $ \ket{t_1} $ as our physical degrees of freedom. Once we have established that a single triplet is localized up to third-order in perturbation, we can easily calculate the spin-gap. For that, we take the unperturbed state excited state $$\ket{\psi_1}=\ket{s}_1\ket{s}_2\cdots\ket{t_1}_i\cdots\ket{s}_{N_d-1}\ket{s}_{N_d}$$ and calculate its energy $e_{1P}$ using perturbation theory (the subscripts of the kets represent a $J$ bond and $N_d$ is the total number of dimers in the system.). The singlet-triplet gap $\Delta=e_{1P}-e_0$, which up to third order in perturbation reads as
\be
\Delta = J-B-\frac{J'^2}{J}-\frac{J'^{3}}{2 J^{2}}+\mathcal{O}(J'^4).
\ee
\section{Two-particle states}
Now we focus on two-particle excited states, which in the unperturbed case is
$$\ket{\psi_1}=\ket{s}_1\ket{s}_2\cdots\ket{t_1}_{i-1}\ket{t_1}_i\ket{s}_{i+1}\cdots\ket{t_1}_{j-1}\ket{t_1}_j\ket{s}_{j+1}\cdots\ket{s}_{N_d-1}\ket{s}_{N_d}.$$ Again, the effect of $\mathcal{H}$ is included via degenerate perturbation~\cite{Sakurai2020}. This allows us to obtain an effective Hamiltonian upon a hard-core boson projection. For $B>0$, we expect that the lowest energy states with magnetization $M \geq 0$ for the whole system consist of the two lowest energy basis states, $|s\rangle$ and $\left|t_{1}\right\rangle$. Therefore, we take the $S^{z}=1$ triplet as a particle (hard-core boson) and the singlet as a vacancy ($\hat{a}_{i}^{\dagger}|s\rangle_{i}=\left|t_{1}\right\rangle_{i}$ with $i$ being the dimer index, and $\hat{a}_i^\dagger$ being the bosonic creation operator. We note that the two-particle excited states come with a potential and a kinetic contribution. We find that the effective hard-core boson Hamiltonian up to two particles reads as
\bea\label{eq:interaction_H_eff}
\begin{aligned}
\mathcal{H}_\text{eff} \approx  e_{0}&+\Delta \sum_{i} \hat{n}_{i}+V_{1} \sum_{\langle i, j\rangle} \hat{n}_{i} \hat{n}_{j}+V_{2} \sum_{( i, j)} \hat{n}_{i} \hat{n}_{j}+V_{3} \sum_{[ i, j]} \hat{n}_{i} \hat{n}_{j}\\
&+ \tau_{1} \sum_{\mathbf{r}}\left(\hat{a}_{A( \mathbf{r}+\mathbf{a}_{1})}^{\dagger} \hat{a}^{}_{A(\mathbf{r})}+\text { h.c. }\right) \hat{n}_{B(\mathbf{r})}\\
&+\tau_{2} \sum_{\mathbf{r}}\left(\hat{a}_{A(\mathbf{r})}^{\dagger} \hat{a}^{}_{C(\mathbf{r})}+h . c .\right)\hat{n}_{B(\mathbf{r})}+\tau_{2} \sum_{\mathbf{r}}\left(\hat{a}_{A(\mathbf{r})}^{\dagger} \hat{a}^{}_{C(\mathbf{r}-\mathbf{a}_{2})}+h . c .\right) \hat{n}_{B(\mathbf{r}-\mathbf{a}_{1})}\\
&+\tau_{3} \sum_{\mathbf{r}}\left(\hat{a}_{A(\mathbf{r})}^{\dagger} \hat{a}^{}_{C(\mathbf{r})}+h . c .\right) \hat{n}^{}_{B(\mathbf{r}-\mathbf{a}_{1})} +\tau_{3} \sum_{\mathbf{r}}\left(\hat{a}_{A(\mathbf{r})}^{\dagger} \hat{a}^{}_{C(\mathbf{r}-\mathbf{a}_{2})}+h . c .\right) \hat{n}^{}_{B, \mathbf{r}} \\
&+\left\{S(B C A), \mathbf{a}_{1} \rightarrow-\mathbf{a}_{1}+\mathbf{a}_{2}, \mathbf{a}_{2} \rightarrow-\mathbf{a}_{1}\right\}+\left\{S(C A B), \mathbf{a}_{1} \rightarrow-\mathbf{a}_{2}, \mathbf{a}_{2} \rightarrow \mathbf{a}_{1}-\mathbf{a}_{2}\right\} \\
\end{aligned}
\eea
where $\hat{n}_{i}=\hat{a}_{i}^{\dagger} \hat{a}_{i}$ is the number operator of hardcore bosons on $i$-th dimer. In the static part of \eqref{eq:interaction_H_eff},
\begin{equation}
\begin{aligned}
V_{1} &= \frac{J'}{2}+\frac{J'^{2}}{2 J}+\O(J'^4) \\
V_{2} &= \frac{J'^{2}}{2 J}+\frac{3 J'^{3}}{4 J^{2}}+\O(J'^4) \\
V_{3} &= \frac{J'^{3}}{8 J^{2}}+\O(J'^4) 
\end{aligned}
\end{equation}
and the position of the triplet pairs $\langle i, j\rangle$, $( i, j)$, and $[ i, j]$ are given by
$$\langle i, j\rangle \in \left\{\begin{array}{l}\left\{A(\mathbf{r}), B\left(\mathbf{r}\right)\right\} \\\left\{A(\mathbf{r}), B\left(\mathbf{r} - \mathbf{a}_{1}\right)\right\} \\
\left\{B(\mathbf{r}), C\left(\mathbf{r}\right)\right\} \\\left\{B(\mathbf{r}), C\left(\mathbf{r} + \mathbf{a}_{1}-\mathbf{a}_{2}\right)\right\} \\
\left\{C(\mathbf{r}), A\left(\mathbf{r}\right)\right\} \\\left\{C(\mathbf{r}), A\left(\mathbf{r} + \mathbf{a}_{2}\right)\right\} \end{array}\right.$$
$$( i, j) \in\left\{\begin{array}{l}\left\{A(\mathbf{r}), A\left(\mathbf{r} \pm \mathbf{a}_{2}\right)\right\} \\ \left\{B(\mathbf{r}), B\left(\mathbf{r} \pm \mathbf{a}_{1}\right)\right\} \\ \left\{C(\mathbf{r}), C\left(\mathbf{r} \pm \mathbf{a}_{1} \mp \mathbf{a}_{2}\right)\right\}\end{array}\right.$$
$$[ i, j] \in\left\{\begin{array}{l}\left\{A(\mathbf{r}), B\left(\mathbf{r}-\mathbf{a}_{1}+\mathbf{a}_{2}\right)\right\} \\ \left\{A(\mathbf{r}), B\left(\mathbf{r}-\mathbf{a}_{2}\right)\right\} \\ \left\{B(\mathbf{r}), C\left(\mathbf{r}+\mathbf{a}_{1}\right)\right\} \\ \left\{B(\mathbf{r}), C\left(\mathbf{r}-\mathbf{a}_{2}\right)\right\} \\ \left\{C(\mathbf{r}), A\left(\mathbf{r}-\mathbf{a}_{1}+\mathbf{a}_{2}\right)\right\} \\ \left\{C(\mathbf{r}), A\left(\mathbf{r}+\mathbf{a}_{1}\right)\right\}\end{array}\right..$$
From this part of the Hamiltonian, one can see that the energy of a two-triplet excitation consists of two contributions: (1) the spin gap energy to create a triplet excitation including the self-energy corrections, and (2) the interaction between the excited triplets. For the kinetic part,
$$
\begin{aligned}
\tau_{1} \approx &\frac{J^{\prime 2}}{4 J}+\frac{3J^{\prime 3}}{8J^{2}}+\O(J^4)\\
\tau_{2} \approx &-\frac{J^{\prime 2}}{4 J}+\frac{J^{\prime 3}}{4J^{2}}+\O(J^4)\\
\tau_{3} \approx &\frac{J^{\prime 2}}{4 J}+\frac{5J^{\prime 3}}{16J^{2}}+\O(J^4)
\end{aligned}
$$
and in the last line $S(\cdots)$ represents the cyclic permutations. Note that unlike single particle hopping, correlated hopping, i.e. the hopping of a triplet assisted by another triplet in its vicinity, occurs in $J'^2$. We also find some particular two particle interactions,
$$
V_{4} \sum_{\left\{ i, j\right\}} n_{i} n_{j},
$$
with
$$
V_{3} \approx \frac{J^{4}}{8 J^{3}}
$$
and
$$
\left\{ i, j\right\}\in\left\{\begin{array}{l}
\left\{A(\mathbf{r}), C\left(\mathbf{r}-\mathbf{a}_{1}+\mathbf{a}_{2}\right)\right\} \\
\left\{A(\mathbf{r}), C\left(\mathbf{r}+\mathbf{a}_{1}-2 \mathbf{a}_{2}\right)\right\} \\
\left\{A(\mathbf{r}), A\left(\mathbf{r} \pm \mathbf{a}_{1} \mp 2 \mathbf{a}_{2}\right)\right\} \\
\left\{B(\mathbf{r}), A\left(\mathbf{r}-\mathbf{a}_{2}\right)\right\} \\
\left\{B(\mathbf{r}), A\left(\mathbf{r}+\mathbf{a}_{1}+\mathbf{a}_{2}\right)\right\} \\
\left\{B(\mathbf{r}), B\left(\mathbf{r} \pm \mathbf{a}_{1} \pm \mathbf{a}_{2}\right)\right\} \\
\left\{C(\mathbf{r}), B\left(\mathbf{r}+\mathbf{a}_{1}\right)\right\} \\
\left\{C(\mathbf{r}), B\left(\mathbf{r}-2 \mathbf{a}_{1}+\mathbf{a}_{2}\right)\right\} \\
\left\{C(\mathbf{r}), C\left(\mathbf{r} \pm 2 \mathbf{a}_{1} \mp \mathbf{a}_{2}\right)\right\}
\end{array}\right.
$$
that only contributes in the fourth order. 

\begin{figure}
\includegraphics[width=0.8\columnwidth]{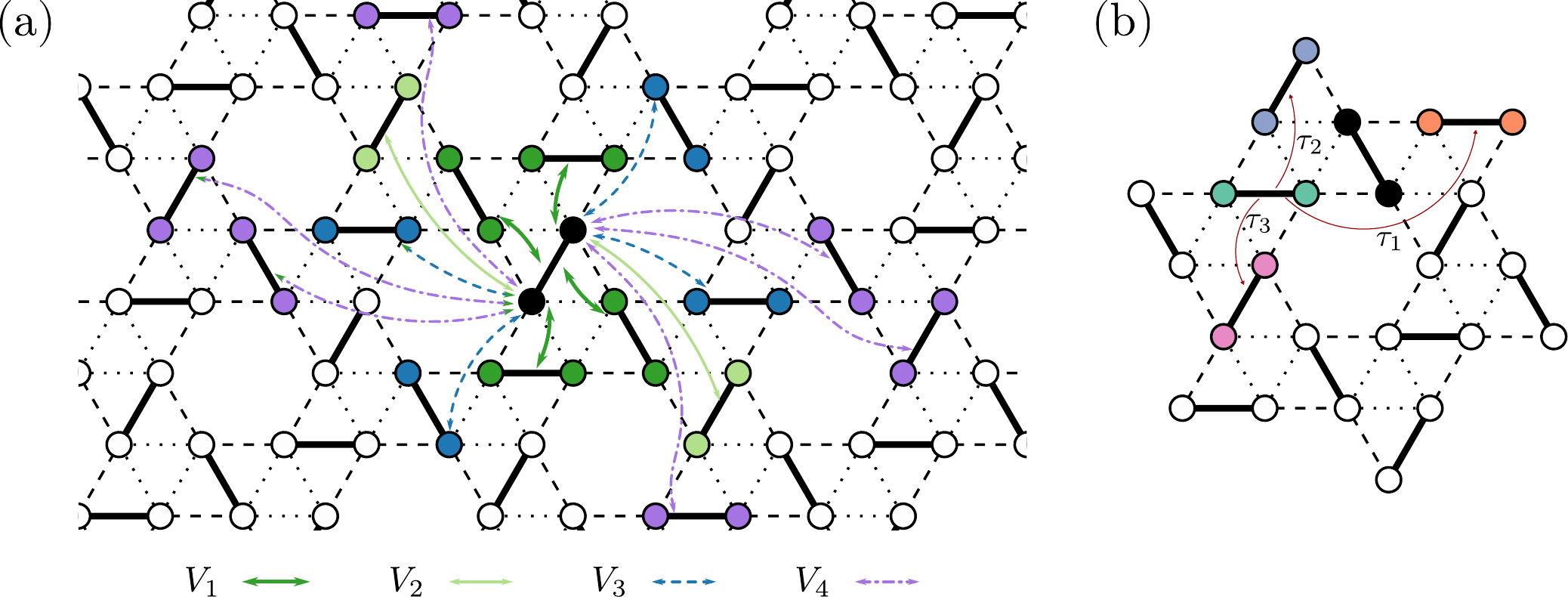}
\caption{(a) Repulsive interaction between a pair of particles. The filled black symbols represent the reference dimer, and all other filled symbols represent a particle that feels repulsion from the reference dimer. (b) Hopping processes of a dimer excitation (colored symbols) near another triplet excitation (filled bonds).} \label{fig-interaction}
\end{figure}

The two-particle interactions and correlated hopping (a hopping of a triplet assisted by the) are graphically represented in Fig.~\ref{fig-interaction}. The interesting difference between SSM and MLM is that the repulsion between two triplets across the diagonal of a square is non-zero in SSM; however, the interaction between two triplets across the long diagonal of a hexagon in MLM is zero up to fourth-order in perturbation.

For $B=0, J_{c r i}^{\prime}$ for SSM is $0.67$ and $J_{c r i}^{\prime}$ for MLM is $0.74$. In the perturbative calculations, the gap closes at $J^{\prime}=0.839$ for $B=0$. This indicates that with increasing $J^\prime$ there must be a first-order transition out of the singlet phase for both models, as a gap closing indicates a second-order transition. This is indeed true for SSM~\cite{Mambrini2006,Corboz2013,Chung2001,deconfined,Farnell2011}.

\section{Magnetization}
Next, we find the magnetization plateaus and the corresponding triplet structures that MLM can produce in the presence of an external magnetic field. With two-particle interactions only between Ising spin-like hard-core bosons, one can do so by employing the standard classical single-flip Monte-Carlo (MC) simulation of the effective Hamiltonian in \eqref{eq:interaction_H_eff} while ignoring the kinetic terms. For that, we take a kagom\'e lattice (the effective lattice of the dimers) with $N_1\times N_2$ unit-cells, i.e. a total of $L=N_1\times N_2\times 3$ dimers with a periodic boundary condition. Here, $N_1$ and $N_2$ denote the number of unit-cells along the lattice vector $\mathbf{a}_1$ and $\mathbf{a}_2$, respectively. We start our MC simulation by taking a random spatial distribution of the hard-core bosons, $\mathbf n = (n_1, \cdots, n_L)$, where $n_i$ can either be $0$ (empty) or $1$ (filled). For our MC simulation, we adopt a simulated annealing algorithm where we start at a high temperature and slowly reduce our temperature while allowing the system to thermalize at each temperature step. 
For each temperature, an increasing number of MC cycles are performed. In each MC update, one would usually choose a random dimer ($r$) and flip its state ($n_r\rightarrow 1-n_r$) and accept the move based on the standard Metropolis algorithm. In a single MC cycle, $L$ individual Metropolis updates are attempted. However, this process becomes highly inefficient in the presence of a magnetic field as the simulation can very easily get stuck in a local energy minimum. Therefore, we also introduce a process where we choose two random sites ($r$ and $r^{\prime}$), and swap their states ($n_{r}\leftrightarrow n_{r^{\prime}}$). The swapping ensures that the energy cost to excite triplets, $\propto \Delta$ ($\sim h$), does not change and the system is allowed to arrange itself solely based on the interactions. We slowly increase the probability of such swaps with every MC cycle, so that our system can exploit this feature at every temperature step. This algorithm is parallelized over $32$ different initial states and we average over the individual results.

In our simulations, we have considered different geometries of the lattice, i.e. different combinations of $N_1$ and $N_2$, the initial temperature being $T_i =2\Delta $ and decreased the temperature to a minimum of $T_f (\sim J^{\prime 4})$ after $N_T = 10^5$ annealing steps. At $T_i$, $10^2$ cycles are performed, which is increased linearly to $10^4$ cycles at $T_{f}$.

\begin{figure*}
\includegraphics[width=\textwidth]{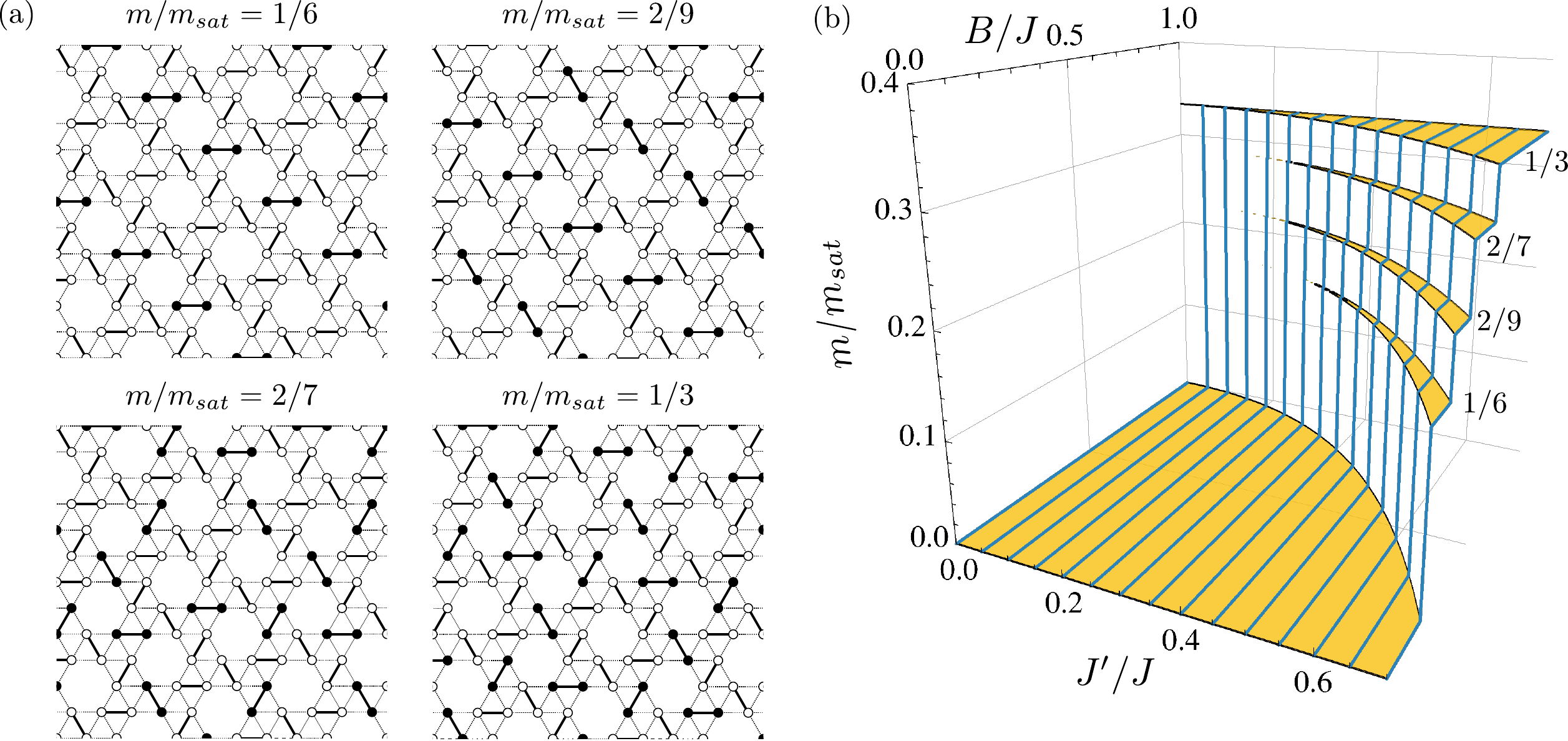}
\caption{Left panel: The structures at the plateaus found from classical Monte Carlo simulations with $\Delta$ given in \eqref{eq:delta_4} and including $V_4$. The triplet excitations (singlets) are shown by filled (open) symbols. Right panel: Magnetization processes of the MLM as a function of $J^\prime$ and $h$.} \label{fig-plateau_analytic}
\end{figure*}

In our MC simulation, we find four stable magnetization plateaus, namely, $2/9$, $2/7$, and $1/3$. We sample single configurations from our MC simulations to check the triplet configuration of the magnetization value, which are shown in Fig.~\ref{fig-plateau_analytic} (a). In the presence of $V_4$ we can predict the appearence of an additional $1/6$ plateau. We estimate the energies of these patterns from \eqref{eq:interaction_H_eff} to find
$$
\begin{aligned}
e_{1/6} &=\frac{1}{6}\Delta\\
e_{2 / 9} &=\frac{2}{9}\left(\Delta+V_4\right)\\
e_{2 / 7} &=\frac{2}{7}\left(\Delta+V_{3}+V_4\right) \\
e_{1 / 3} &=\frac{1}{3}\left(\Delta+2 V_{3}+V_4\right).
\end{aligned}
$$
By comparing the energies above we obtain the magnetization phase diagram shown in Fig.~\ref{fig-plateau_analytic} (b), which is in full agreement with our MC simulation.

\section{Effect of single-particle kinematics}
\begin{figure}
\includegraphics[width=0.5\columnwidth]{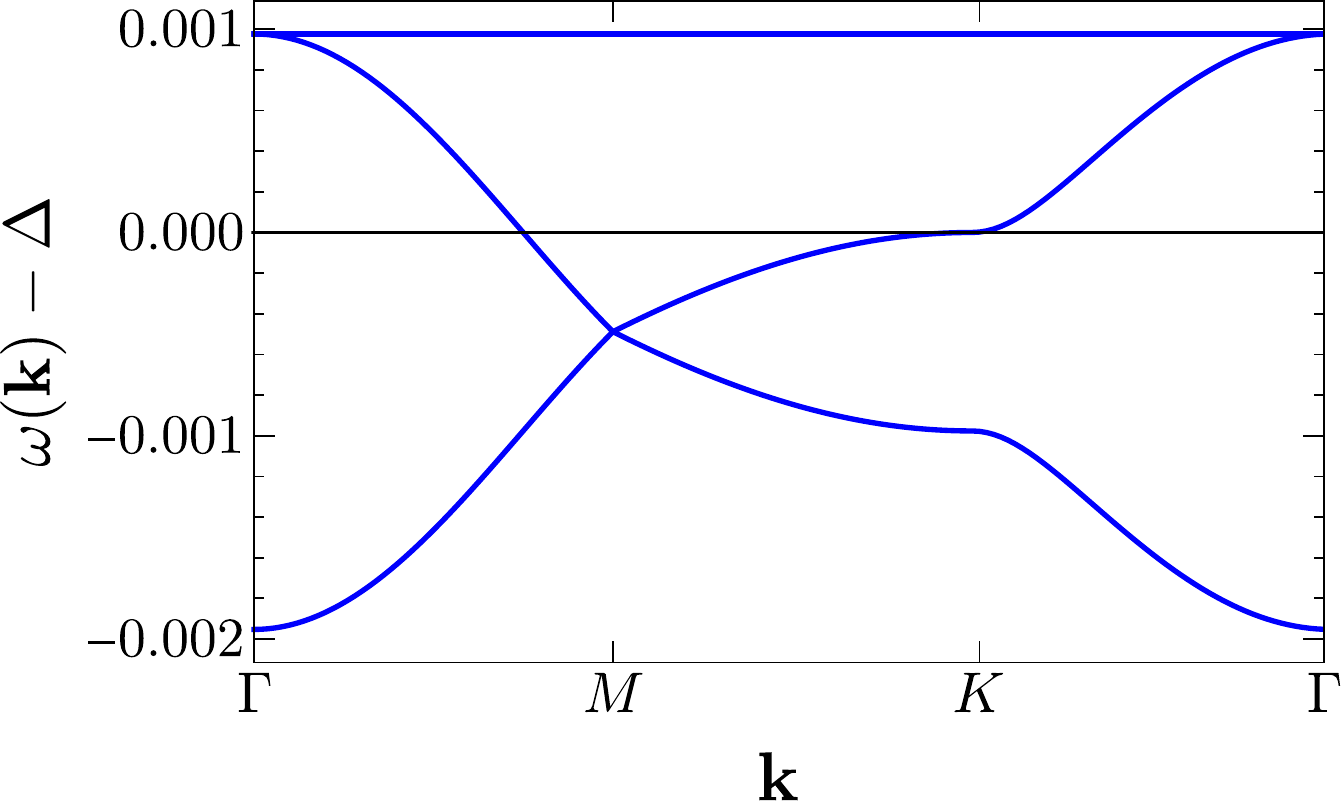}
\caption{(a) The bandstructure of the effective tight-binding Hamiltonian $\mathcal{H}_\text{eff}^0=\hat{T}_{1p}+\Delta_0\sum_i \hat{n}_i$. Note that this tight-binding model is defined on the kagome lattice. The minimum of the bandstructure occurs at $\mathbf k=\Gamma$ which renormalizes the singlet-triplet gap to $\Delta$ given in \eqref{eq:delta_4}.} \label{fig-1p-band}
\end{figure}
The single triplet hopping starts at fourth order in perturbation where a triplet can hop to the neighbouring dimers. This will add one more kinetic term to our effective Hamiltonian in~\eqref{eq:interaction_H_eff}:
\bea
\hat{T}_{1p}&=&\tau_{0} \sum_{\mathbf{r}}\left(\hat{a}_{B( \mathbf{r})}^{\dagger} \hat{a}^{}_{A(\mathbf{r})}+\hat{a}_{C( \mathbf{r})}^{\dagger} \hat{a}^{}_{A(\mathbf{r})}+\hat{a}_{B( \mathbf{r}-\mathbf a_1)}^{\dagger} \hat{a}^{}_{A(\mathbf{r})}+\hat{a}_{C( \mathbf{r}-\mathbf a_2)}^{\dagger} \hat{a}^{}_{A(\mathbf{r})}\right.\\
&+&\left.\hat{a}_{A( \mathbf{r}+\mathbf a_1)}^{\dagger} \hat{a}^{}_{B(\mathbf{r})}+\hat{a}_{C( \mathbf{r}+\mathbf a_1-\mathbf a_2)}^{\dagger} \hat{a}^{}_{B(\mathbf{r})}+\hat{a}_{A( \mathbf{r}+\mathbf a_2)}^{\dagger} \hat{a}^{}_{C(\mathbf{r})}+\hat{a}_{B( \mathbf{r}-\mathbf a_1+\mathbf a_2)}^{\dagger} \hat{a}^{}_{C(\mathbf{r})}+\text { h.c. }\right)
\eea
where $\tau_0=-J'^4/8J^3$. This kinetic term is identical to a bosonic tight-binding model on the kagome lattice. With the chemical potential term, $\Delta_0 \sum_{i} \hat{n}_{i}$, we obtain the well-known kagome band-structure with minimum at $\mathbf k=\Gamma$ (Fig. \ref{fig-1p-band}). $$\Delta_0=J - \frac{J'^2}{J} - \frac{J'^3}{2 J^2} - \frac{J'^4}{8 J^3}+\mathcal{O}(J'^5)$$ is the singlet-triplet gap assuming dispersionless triplets. Therefore, at fourth order of perturbation the true energy gap, $\Delta$, includes the contribution from single particle dispersion and becomes 
\be\label{eq:delta_4}
\Delta=J - \frac{J'^2}{J} - \frac{J'^3}{2 J^2} - \frac{5J'^4}{8 J^3}+\mathcal{O}(J'^5).
\ee
\section{Multi-particle (two and three) Bound States}
\begin{figure}
\includegraphics[width=\columnwidth]{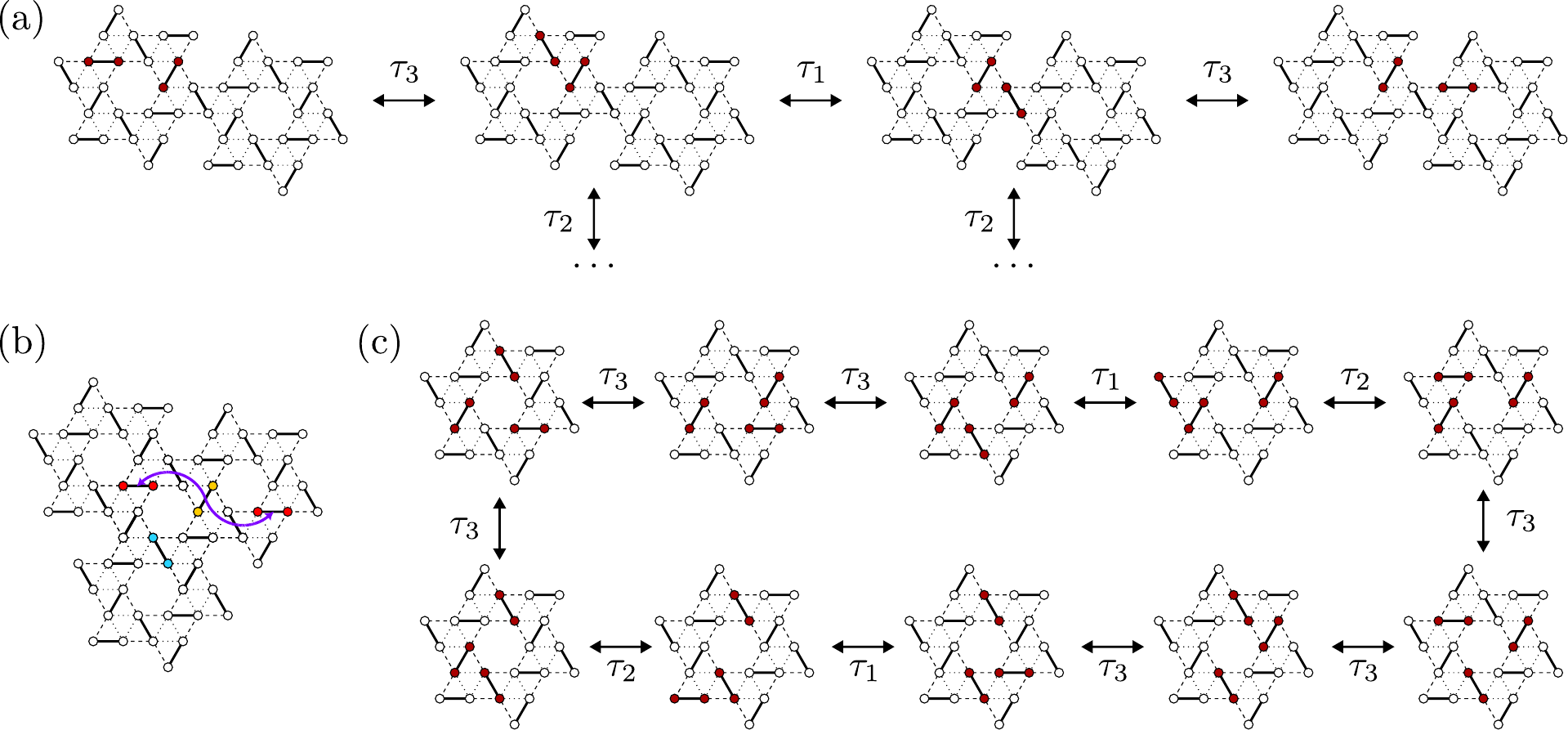}
\caption{(a) Four bases of a bound state of two dimer triplets. There are $8$ other basis states which are symmetry related to the states shown here. (b) A representative three-particle process, where two particles perform a correlated hopping, while one stays put. The red, cyan, and yellow dimers represent the starting configuration of the triplets. The red dimer performs a correlated hopping with respect to the yellow dimer. (c) The representative correlated the three-particle hopping process, which can lead to three-particle bound states.} \label{fig-2p-3p}
\end{figure}
For this part of our calculations, we restrict ourselves upto third-order perturbative corrections. When two triplets are distant from each other, their mobility is severely limited, and the energy they acquire while moving on a lattice is comparatively negligible. On the other hand, when the two are adjacent to each other, the situation is completely different, where a correlated hopping makes a coherent motion of two triplets possible, exploiting which a pair of triplets can form a bound state. Obviously, when we place two triplets next to each other, the repulsive interaction between them is linear in $J^{\prime}$. However, they can hop and move away from each other to a state which has an interaction, $V_2^{\prime}$ which is only cubic in $J^{\prime}$. The hopping element that facilitates this move is $\propto J^{\prime 2}$. Thus, the two triplets can gain enough kinetic energy to form a two-particle bound state. The energy of the bound state is found by diagonalizing the hopping matrix in the appropriate basis. We find that all unique (up to an arbitrary lattice translation) two-particle states can be written in a 12-dimensional basis $|i\otimes j\rangle$, where $i$ and $j$ denote occupied dimers, given by
\begin{equation}
|i\otimes j\rangle \in \left\{\begin{array}{l}
\left| A(\mathbf r)\otimes B(\mathbf r) \right\rangle \\
\left| A(\mathbf r)\otimes C(\mathbf r) \right\rangle \\
\left| B(\mathbf r)\otimes C(\mathbf r) \right\rangle \\
\left| A(\mathbf r + \mathbf a_2)\otimes B(\mathbf r + \mathbf a_2 - \mathbf a_1) \right\rangle \\
\left| C(\mathbf r)\otimes A(\mathbf r + \mathbf a_2) \right\rangle \\
\left| C(\mathbf r)\otimes B(\mathbf r + \mathbf a_2 - \mathbf a_1) \right\rangle \\
\left| C(\mathbf r)\otimes, A(\mathbf r + \mathbf a_2 - \mathbf a_1) \right\rangle \\
\left| C(\mathbf r)\otimes A(\mathbf r + \mathbf a_1) \right\rangle \\
\left| A(\mathbf r)\otimes B(\mathbf r + \mathbf a_2 - \mathbf a_1) \right\rangle \\
\left| B(\mathbf r)\otimes A(\mathbf r + \mathbf a_2) \right\rangle \\
\left| B(\mathbf r)\otimes C(\mathbf r - \mathbf a_2) \right\rangle \\
\left| C(\mathbf r + \mathbf a_2)\otimes B(\mathbf r + \mathbf a_2 - \mathbf a_1) \right\rangle \\
\end{array}\right.
\end{equation}
Four of these basis states are shown in Fig.~\ref{fig-2p-3p} (a), the rest are related to these four via a lattice rotation. The matrix representing the correlated hopping matrix in the above basis is given by
\begin{align}
T^{2P}_{\mathbf k} =
\left(
\begin{array}{c|c}
\begin{array}{cccccc}
2 \Delta + V_{1} & \tau_{2} & \tau_{2} & 0 & 0 & \tau_{1} e^{- i \phi_{2, \mathbf k}}\\
\tau_{2} & 2 \Delta + V_{1} & \tau_{2} & 0 & \tau_{1} e^{i \phi_{1, \mathbf k}} & 0 \\
\tau_{2} & \tau_{2} & 2 \Delta + V_{1} & \tau_{1} & 0 & 0\\
0 & 0 & \tau_{1} & 2 \Delta + V_{1} & \tau_{2} & \tau_{2} \\
0 & \tau_{1} e^{- i \phi_{1, \mathbf k}} & 0 & \tau_{2} & 2 \Delta + V_{1} & \tau_{2} \\
\tau_{1} e^{i \phi_{2, \mathbf k}} & 0 & 0 & \tau_{2} & \tau_{2} & 2 \Delta + V_{1} \\
\end{array} & \tau_{3} \, \mathbb{I}_{6 \times 6} \\
\hline \\
\tau_{3} \, \mathbb{I}_{6 \times 6} & (2 \Delta + V_{2}) \, \mathbb{I}_{6 \times 6}
\end{array}
\right)
\end{align}
with $\mathbb{I}_{6 \times 6}$ being the identity matrix of dimension $6$, $\phi_{1,\mathbf k} = \mathbf k \cdot \mathbf a_2$, and $\phi_{2,\mathbf k} = \mathbf k \cdot \mathbf a_1$.

Since strong repulsion $V_1$ acts for a pair on adjacent bonds, one may naively expect that such bound motions are not energetically favorable. However, we can take an optimal linear
combination of the states, which avoids the effect of $V_1$ repulsion and gain kinetic energy via coherent motion (see example in Fig.~\ref{fig-2p-3p} (a)). The binding energy, $e_{2P}-2\Delta$, for such a bound state is found to be $-3J^{\prime4}/16J^3$ (lowest order) per particle. Note that this two-particle bound state is distinctively different from SSM.

Drawing inspiration from the two-particle bound state of SSM, we start from a situation as depicted in Fig.~\ref{fig-2p-3p} (b). Firstly, one can see that any two particles can make a coherent motion encountering no effect from the third one. Note that apart from the initial state, in the other states, the third particle does not interact with the particle that is hopping. Therefore, there exists three independent two-particle hopping in such a three-particle state. Secondly, the three particles themselves can perform a coherent motion around a hexagon and form a three-particle bound state. In this truncated three particle basis shown in Fig.~\ref{fig-2p-3p} (c) the binding energy is found to be $J^{\prime3}/24J^2-J^{\prime4}/8J^3$ per particle. We find that the binding energy is positive for small $J^\prime$. However, due to the extensive size of the full three-particle Hilbert space, we could not perform the three-particle calculations with its full complexity and analytically evaluate the binding energy of the three particle states. It seems plausible that a more elaborate evolution of three particle processes might make the cubic contribution to the binding energy could disappear or even change sign and produce stable three-particle bound states, at least for larger $J^\prime$.

\begin{figure}
\includegraphics[width=0.75\columnwidth]{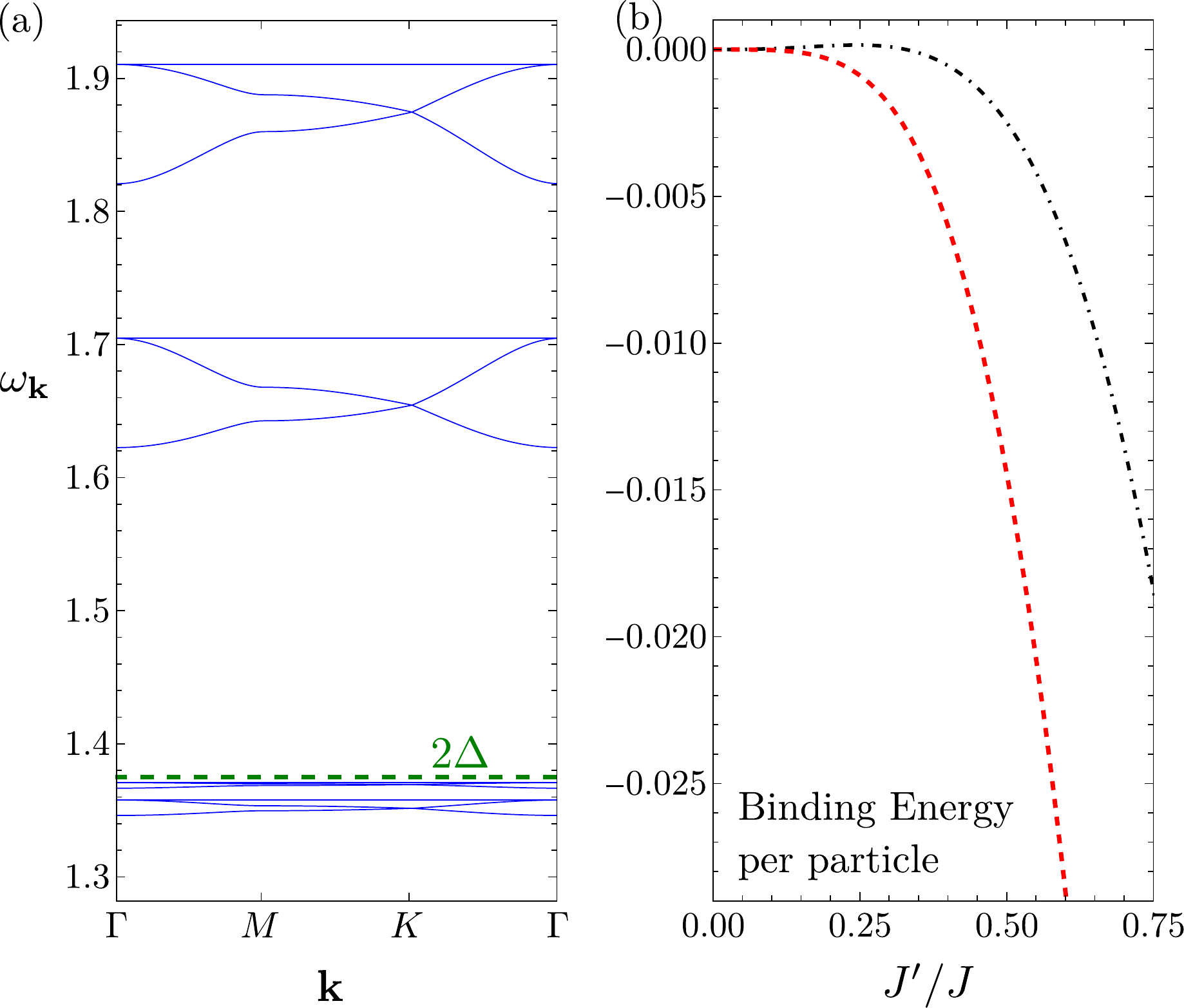}
\caption{(a) Energy bands of the effective Hamiltonian up to
third order of $J^\prime/J=0.5$. A two-particle bound state consists of six branches, which is apparently lower than the two-particle threshold, shown by the green dashed line. (b) The per-particle binding energy of two- (red, dashed) and three-particle (black, dash-dotted) bound states as a function of $J^\prime/J$.} \label{fig-bound}
\end{figure}

\bibliography{Refs}